# Game Data Mining Competition on Churn Prediction and Survival Analysis using Commercial Game Log Data

EunJo Lee, Yoonjae Jang, DuMim Yoon, JiHoon Jeon, Seong-il Yang, Sang-Kwang Lee, Dae-Wook Kim, Pei Pei Chen, Anna Guitart, Paul Bertens, África Periáñez, Fabian Hadiji, Marc Müller, Youngjun Joo, Jiyeon Lee, Inchon Hwang and Kyung-Joong Kim

*Abstract—* Game companies avoid sharing their game data with external researchers. Only a few research groups have been granted limited access to game data so far. The reluctance of these companies to make data publicly available limits the wide use and development of data mining techniques and artificial intelligence research specific to the game industry. In this work, we developed and implemented an international competition on game data mining using commercial game log data from one of the major game companies in South Korea: NCSOFT. Our approach enabled researchers to develop and apply state-of-the-art data mining techniques to game log data by making the data open. For the competition, data were collected from Blade & Soul, an action role-playing game, from NCSOFT. The data comprised approximately 100 GB of game logs from 10,000 players. The main aim of the competition was to predict whether a player would churn and when the player would churn during two periods between which the business model was changed to a free-to-play model from a monthly subscription. The results of the competition revealed that highly ranked competitors used deep learning, tree boosting, and linear regression.

## I. Introduction

Game artificial intelligence (AI) competition platforms help researchers access well-defined benchmarking problems to evaluate different algorithms, test new approaches, and educate students [1]. Since the early 2000s, considerable effort has focused on designing and running new game AI competitions using mathematical, board, video, and physical games [2]. Despite a few exceptions, most of the research has concentrated on building AI players to play challenging games such as StarCraft and simulated car racing and fighting games; these competitions commonly rank AI players based on the results of numerous game plays, using, for example, final scores and win ratios. Recently, there have been special competitions that target human-likeness [3], general game playing [4], and the learning ability of AI players. However, there have been few competitions on content creation [5], game player modeling, or game data mining.

There have been many attempts to analyze game players. Bartle proposed analyzing multi-user dungeon game players (MUDs) into four types in 1996 [13] and expanded the model to eight types [14]. Quantic Foundry proposed six gamer motivation models based on 12 motivation factors from 5-min surveys with 250,000 gamers [15]. Others have attempted to model game players based on data analysis [16][17][18][19][20]. In recent years, game data mining has become increasingly popular. Data mining techniques extract useful information from large databases and are widely adopted in practical data analysis [21][22].

Game companies generate a large amount of game player data based on their actions, progress, and purchases. From such data, it is possible to model the users' patterns [6] and attain useful information, including in-game dynamics, a user's likelihood of churning, lifetime, and user clusters. For example, Kim et al. discovered a "real money trading (RMT)" pattern from the MMORPG Lineage game [8] and detected Bots and socio-economy patterns from the MMORPG Aion game [7][9]. Despite its potential benefits to the game industry, collaboration between academics and game companies to develop and apply advanced techniques to game log data has not flourished to the extent of AI game player development or content generation. Figure 1 shows

K.-J. Kim, D.-M. Yoon, and J-H. Jeon are with the Computer Science and Engineering Department, Sejong University, Korea (e-mail: kimkj@sejong.ac.kr, corresponding author)

S.-I. Yang, S.-K. Lee and D.-W. Kim are with ETRI (Electronics and Telecommunications Research Institute) Korea (e-mail: siyang@etri.re.kr)

E.-J. Lee and Y.-J. Jang are with NCSOFT, Korea (e-mail: gimmesilver@ncsoft.com)

P.P. Chen, A. Guitart, P. Bertens, and Á. Periáñez are with Game Data Science Department, Yokozuna Data, Silicon Studio, 1-21-3 Ebisu Shibuya-ku, Tokyo, Japan (e-mail: {peipei.chen, anna.guitart, paul.bertens, africa.perianez} @siliconstudio.co.jp)

Fabian Hadiji and Marc Müller are with goedle.io GmbH (e-mail: {fabian, marc}@goedle.io)

Youngjun Joo, Jiyeon Lee, and Inchon Hwang are with Dept. of Computer Engineering, Yonsei University, Korea (e-mail: {chrisjoo12, jitamin21, ich0103}@gmail.com)

that the number of studies on game data mining plateaued after a sudden increase in 2013.

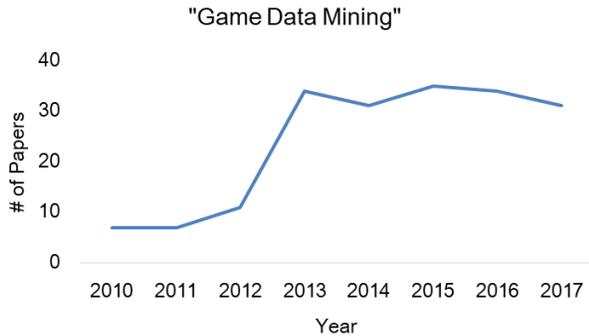

*Figure 1. Number of papers on game data mining in 2010–2017 (from Google Scholar using "Game Data Mining" keywords)*

The purpose of our game data mining competition was to promote the research of game data mining by providing commercial game logs to the public. In coordination with NCSOFT, one of the largest game companies in South Korea, approximately 100 GB of game log data from Blade & Soul were made available.

We hosted the competition for five months from March 28, 2017, to August 25, 2017. During this period, we had 300 registrations on the competition's Google Groups that were given access to the log data. Finally, we received 13 submissions for Track 1 (churn prediction) and 5 submissions for Track 2 (survival analysis). The participants predicted the future behavior of users by applying different sets of state-of-the-art techniques such as deep learning, ensemble tree classifiers, and logistic regression.

The contributions of the game data mining competition are listed below.

- The competition opened commercial game log data from an active game to the public for benchmarking purposes. After the competition, NCSOFT allowed users to copy and retain data for educational, scholarly, and research applications. In addition, labels for the test dataset were made publicly available.
- Similar to the ImageNet competition, the competition provided a test server for participants to submit their intermediate performance and benchmark predictions using 10% of the test data.
- The competition problems applied practical containments. For example, a much longer time span (e.g., three weeks) was given between the training data and prediction window, reflecting the minimum time required to develop and execute churn prevention strategies to retain potential churners. This enhanced the difficulty of the competition problem compared with the conventional time span of one to two weeks.
- The competition was designed to incorporate concept drift, specifically, a change in the business model, to measure the robustness of the participants' models when applied to constantly evolving conditions. Consequently, the competition comprised two test datasets, each from different periods. Between the two periods, the aforementioned business model change took place. The final standings of entries were determined based on the harmonic average of final scores from both test datasets.

II. COMPETITION PROBLEMS: CHALLENGES IN GAME DATA MINING COMPETITION

Blade & Soul, launched in June 2012, features a combination of epic martial-arts actions with highly customizable characters [1] (Figure 2). Along with the solo-play experience with numerous quests and dungeons, Blade & Soul incorporates party dungeons and quests, as well as group combination attacks cultivating true camaraderie among party members. Additionally, fighting in an intense player-versus-player arena mode has allowed the game to gain global popularity, as evidenced by the participation of hundreds of teams from nine regions in the most recent 2017 World Championship held in Seoul, South Korea.

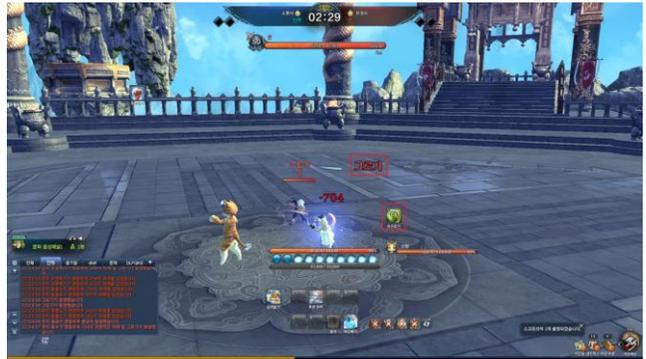

*Figure 2. Screenshot from Blade & Soul*

In this study, the competition consisted of two tracks. Participants could choose to participate in one or both of these tracks. Track 1 aimed to predict whether target users had churned, and Track 2 required participants to predict the survival time of target users. Both tracks used the same dataset consisting of one training set and two test sets. Each set comprised in-game activity logs of sampled players from different periods. To maximize the benefits of the churn prevention model, the competition problem was structured using data from an actively serviced game, Blade & Soul, and considered the following factors.

---

[1] http://www.bladeandsoul.com/en/

*Table 1. Summary of churn prediction study*

| Year | References | Games (Publisher) | Genre (Platform) | Payment System | Target Customers | Churn Definition (Inactivity Period) | Prediction Point |
|---|---|---|---|---|---|---|---|
| 2017 | GDMC 2017 Competition | Blade & Soul (NCSOFT) | MMORPG (PC) | Monthly Charge, Free-to-Play | Highly Loyal Users (cumulative purchase above the certain threshold) | five weeks | Churn after three weeks |
|  |  |  |  |  |  |  | Survival Analysis |
| 2017 | Kim *et al*.,[12] | Dodge the Mud, Undisclosed, TagPro | Causal Game (Online/Mobile) | Free-to-Play | All gamers | ten days | Churn within 10 days |
| 2016 | Tamassia *et al.*, [26] | Destiny (Bungie) | MMOG (Online) | Package Price | Randomly sampled players (play time > 2 hours) | four weeks | Churn after four weeks |
| 2016 | Perianez *et al*. [11] | Age of Ishtaria (Silicon Studio) | Social RPG (Mobile) | Free-to-Play | High value players (whales) | ten days | Survival Analysis |
| 2014 | J. Runge *et al.*, [10] | Diamond Dash and Monster World Flash (Wooga) | Social Game (Mobile) | Free-to-Play | High Value Player (top 10% of all paying players) | two weeks | Churn within the Week |

## A. Prediction targets

To benefit the most from churn prediction and prevention, prediction targets should be those that provide the most profit if retained. Naturally, not all users provide game companies with the same profit; in fact, most users are casual game players accounting for a small proportion of sales, and there are even users who undermine the game services [23]. These light, casual, and malicious players were excluded from the scope of this competition, as our focus was on predicting the churn of loyal users only, namely, highly loyal users with a cumulative purchase above a certain threshold and many in-game activities. Given that highly loyal users seldom churn or churn on occasion due to external factors, we expected that the participants' churn prediction performance would not be comparable to prior churn prediction work. Table 1 shows that several previous works included all or most game player types in their churn prediction.

## B. Definition of player churn

Unlike telecommunication services in which user churn can be easily defined and identified by the user unsubscribing [24], such is not the case for online game services. Online game players seldom delete their accounts or unsubscribe, although they have no intention of resuming game play. In fact, according to our analysis, only 1% of the players inactive for over one year explicitly leave the service by deleting their accounts.

Consequently, we decided player churn using consecutive periods of inactivity. However, what length of inactivity should be considered as churn? This is a difficult question to answer, given that there are various reasons behind a player's inactivity. For example, some players may be inactive for several days because they only play on the weekends. Some may appear inactive for a few weeks because they went on a trip or because they had an important exam that month. If the period to decide player churn is too short, the misclassification rate will be high. If the period is too long, on the other hand, the misclassification rate would be lower but it would take longer to determine whether a player had churned or not; consequently, by the time a player is identified as a churner, there would not be sufficient time to persuade him or her to return.

To resolve this dilemma, we defined a churner as a user who does not play the game for more than five weeks. Figure 3 shows the weekly and daily play patterns of players in concordance with their life patterns and weekly server maintenance every Wednesday morning. Thus, in our analysis, a week is defined from one Wednesday to the next Wednesday. Table 1 shows that previous works used ten days or two weeks for the decision of churn. However, in a recent work for the AAA title, Destiny used four weeks for the churn decision, similar to our definition.

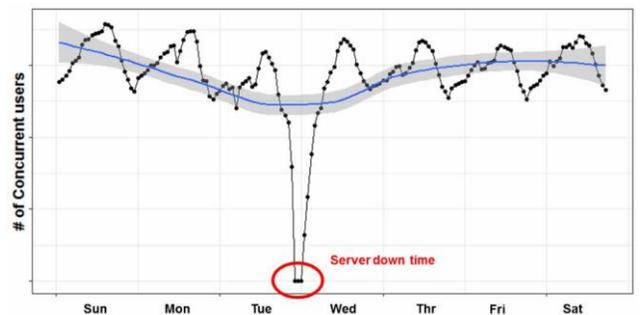

*Figure 3. Time series of concurrent users for one week. Daily and weekly cyclic patterns are shown*

## C. Prediction point

It is likely that user behaviors leading up to the point of churn (e.g., deleting the game character) would play a vital role in churn prediction. However, predicting churn just before the point of leaving would be pointless, as at that point not much can be done to persuade users to stay loyal. Predictions should be made in a timely manner before the churning point so that a churn prevention strategy, namely, a new promotion campaign, content update, and so on, can be implemented to retain potential churners. If the prediction is made too early, the result will not be sufficiently accurate, whereas if the prediction is made too late, not much can be done to retain the user even if the prediction is correct. For our competition, making predictions three weeks ahead of the point of actual churn was deemed effective. Consequently, as shown in Figure 4, we examined user data accounts up to three weeks before the initiation of the five-week churning window.

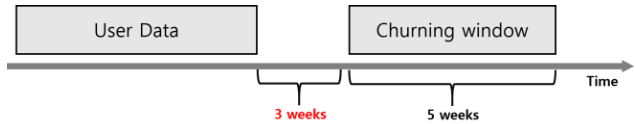

*Figure 4. User data and churning window were used to predict and determine churn, respectively*

### D. Survival analysis

While churn prediction itself is well worthwhile, predicting the specific churn point would increase the value of the model, so this is the second focus of the competition. In addition to churn prediction, we asked participants to perform a survival analysis to predict the survival time. The survival time was then added to the labeled data for the training set. Survival time is defined as the period between the date of last activity from the provided data and the date of most recent activity, which was not provided but instead calculated when all predictions were submitted. Because we can only check for observable periods, this survival analysis is regarded as a censoring problem. We added a '+' sign after survival periods for right-censoring data to distinguish censoring data from churning data.

### E. Concept drift

Various operational issues arise when applying the prediction model to actively serviced and continuously evolving games; concept drift epitomizes how such change can affect a prediction model [25]. Compared to a model that had high performance during the modeling period but showed declining performance over time, a robust model that maintains performance, despite externalities, often proves to be more valuable considering update and maintenance costs, even if the prediction performance of the robust model during the modeling period was not as accurate. To encourage the participants to generate a model that is robust enough to withstand changing conditions over time, we created two test sets each with data from different periods to evaluate model performance. The first test set consisted of data from the period two months after the training data period, and the second consisted of data 7 months after the training data period.

Between the first and second test set periods, there was a significant change in the business model of Blade & Soul, as its subscription-based model was changed to a free-to-play model in December 2016. The chronological positioning of training and test sets and the business model change are shown in Figure 5. In an effort to encourage the participants to form a robust model addressing business model change, we placed no restrictions on using test sets (without data labels) for training.

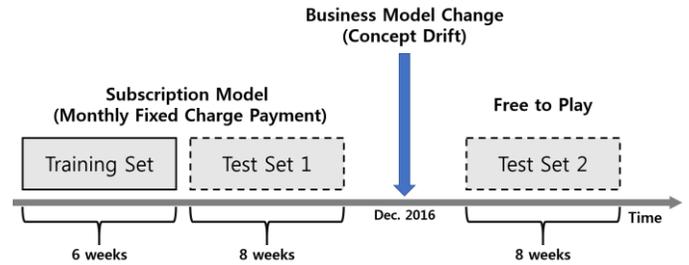

*Figure 5. Test sets 1 and 2 were constructed with data from different periods to reflect the business model change*

## III. RUNNING GAME DATA MINING COMPETITION

### A. Dataset preparation

The competition consisted of two tracks. Three datasets were provided: a training set and two testing sets. Participants were asked to perform churn prediction (Track 1) and survival time prediction (Track 2). Table 2 summarizes the basic dataset information.

*Table 2. General information of the provided dataset*

| Data Set | Time Period | Weeks | Number of Gamers | Data Size* |
|---|---|---|---|---|
| Training | APR-1-2016 ~ MAY-11-2016 | 6 | 4000 (30% churn) | 48G (175m Events) |
| Test Set 1 | JULY-27-2016 ~ SEP-21-2016 | 8 | 3000 (30% churn) | 30G |
| Test Set 2 | DEC-14-2016 ~ FEB-08-2017 | 8 | 3000 (30% churn) | 30G |

For the competition, participants were given access to raw log data of actual Blade & Soul users that captured all of their in-game activities. Among hundreds of log types, 82 main log types, which captured information on connection, character, item, skills, quest, and guild, were used to create the competition dataset. Each log type consisted of 77 fields categorized into common, actor, object, and target fields, as described below.

- Common fields captured information common to all log types such as log type, log creation time, and in-game location of the corresponding action
- Actor fields depicted information regarding the actor of the action with fields such as actor identification (ID), actor level, and actor race.
- Object fields included information on the object of the action, ranging from item ID, item grade, and item quantity to skill name.
- Target fields contained information regarding the target of the action, such as the target character ID, gold, and damage received by the target.

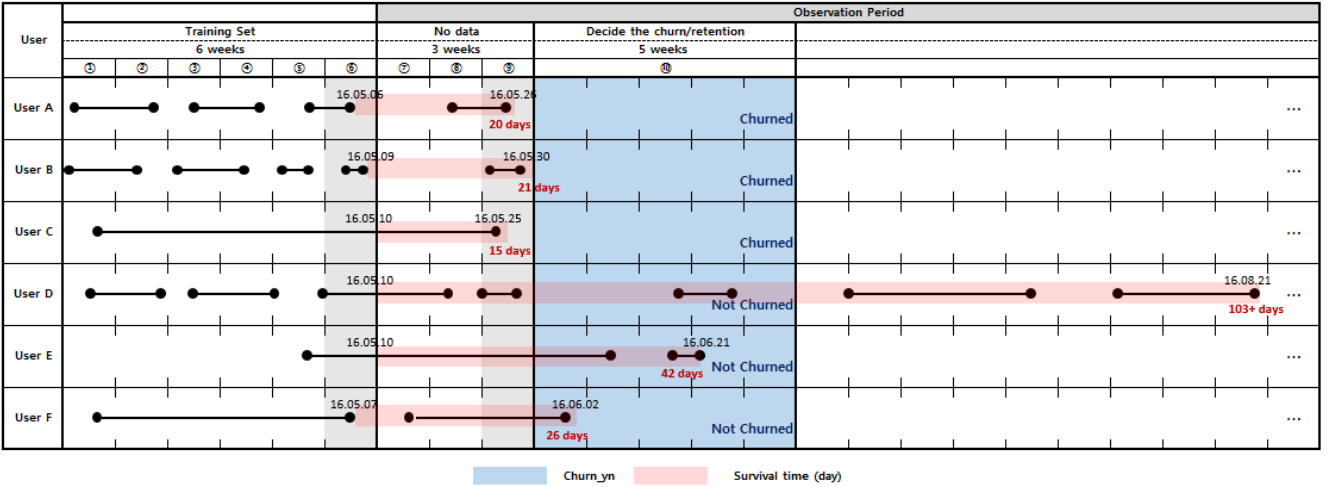

*Figure 6. User activities, churning decision, and survival period*

Each field held different information depending on the log type; detailed log schema and descriptions were provided to the participants, along with the data, through the competition website[2].

As mentioned in the competition problem section, all predictions were made three weeks ahead of the five-week-long churning window, during which users' activities were observed to determine whether they had churned. User churn was determined strictly during the five-week churning window; activities during the three weeks of the "no data" period had no impact on determining user status. Despite activities over other periods, if a user did not exhibit activity during the churning window, the user was considered to have churned. Several examples of users' activities and corresponding churning decisions are shown in Figure 6.

For the second track of the competition, which required participants to predict the survival time of users, any prediction of survival time longer than the actual observed time was considered to be correct given the censoring nature of the problem. For example, if a user's survival time was 103 days by the time submission date, and a participant predicted the survival time of the user to be 110 days, such a prediction was accepted as the correct prediction.

*B. The definition of loyal users*

NCSOFT is constantly detecting malicious users who have an adverse effect on game services. The detection system consists of two steps: One is the step of restricting game service because of collecting concrete evidence for malicious activity, and the other step is the step of follow-up observation because it is a circumstantial suspicion. We excluded the users, who have circumstantial evidence for malicious activities, for the dataset. We note that the detailed rules for the detection cannot be disclosed here, since the information is confidential.

Besides, NCSOFT assigns all users into fourteen grades – the lower grade means a higher loyalty – for customer relationship management. Loyal grades are determined using k-means clustering and take info features such as payment amount, play time and in-game contents usage rate.

Based on this information, we have selected only users who have been assigned at least grade 9 more than once in the last three months. The proportion of the selected group is less than 30% of the total users.

*C. Performance measure definition*

Participants' performances were measured using the average of the F1 score and root mean squared logarithmic error (RMSLE) on the two test sets for Track 1 and Track 2, respectively [Eq. (1) and (2), respectively]:

$$F_1 = 2 \cdot \frac{1}{\frac{1}{recall}+\frac{1}{precision}} = 2 \cdot \frac{precision \cdot recall}{precision+recall} \quad (1)$$

$$\epsilon = \sqrt{\frac{1}{n}\sum_{i=1}^{n}(\log(p_i+1) - \log(a_i+1))^2} \quad (2)$$

where $p_i$ and $a_i$ are the predicted and actual values of the $i$th data in the test set, respectively.

*D. Participants*

On April 15, 2017, sample data and the data schema, as well as competition details, were announced. The entire dataset was made available on April 30, 2017, and the final submission date, originally July 31, 2017, was postponed to August 10, 2017. Starting on May 19, all participants were able to access a test server to validate their work using 10% of the test data.

For efficient communication with the participants, we opened a Google Groups page[3] in which participants were required to register for access to the log data. As shown in Figure 7, the number of registrants who joined the Google Groups increased steadily every month from the end of March 2017 when the group was created. For the competition, we received 13 submissions from South Korea, Germany, Finland, and Japan for Track 1 and 5 submissions for Track 2. The details of all participants are summarized in the supplementary materials[4].

---

[2] https://cilab.sejong.ac.kr/gdmc2017/

[3] https://groups.google.com/d/forum/gdmc2017

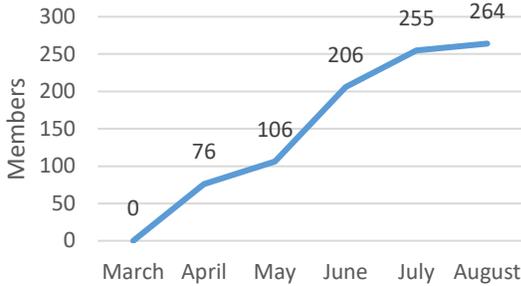

*Figure 7. Number of registrants in the Google Groups site*

## IV. COMPETITION RESULTS

### A. Track 1

This track aimed to predict the churn of the gamer. A total of 13 teams participated in this track. The Yokozuna Data team won, with a final total score of approximately 0.62. To facilitate an easy start and provide a reference point, we provided a tutorial with a baseline model on the competition page[5]. In the tutorial, we used 22 simple count-based features and lasso regression. An F1 score of 0.48 was achieved, indicating that the eight submissions among the 13 entries outperformed the baseline model.

The winning performance with an F1 score of 0.62 was similar to the predictability of models introduced by previous churn prediction work [27]. Considering that there were two constraints that greatly hindered prediction accuracy: targeting loyal users only and predicting based on data from specific time periods, the fact that the winning team showed comparable or even better results underscores the novelty of the work provided by the participants.

Interestingly, participants with lower ranks generally performed better on the first test set than on the second set, reflecting data after the change of business model from a monthly subscription model to a free-to-play model. On the other hand, those with higher ranks performed better on predicting user churn during the second set.

The final standing differed significantly from the ranking on the test server, as shown in Table 3. We believe such a difference in standing is due to the fact that the test server measured performance using only 10% of the test data, whereas the final performance measurement was conducted with the remaining 90% of the test data. Consequently, we suspect that many models that did not perform relatively well suffered from overfitting to the test-server results. In contrast, the Yokozuna Data team who won the competition specifically attempted to avoid overfitting using features with the same distribution throughout a different test set. Given its superior performance, we believe that Yokozuna Data invented a robust model that was properly generalized.

*Table 3. Competition results (the arrows indicate the change of ranks from the test server)*

*(a) Track 1*

| Rank | Team | | Precision | Recall | F1 score | Final score |
|---|---|---|---|---|---|---|
| 1 | **Yokozuna Data** (▲3) | **Test1** | **0.55** | **0.69** | **0.61** | **0.62** |
| | | **Test2** | **0.54** | **0.76** | **0.63** | |
| 2 | UTU (▲1) | Test1 | 0.53 | 0.71 | 0.60 | 0.60 |
| | | Test2 | 0.60 | 0.60 | 0.60 | |
| 3 | TripleS (▼1) | Test1 | 0.54 | 0.62 | 0.57 | 0.60 |
| | | Test2 | 0.56 | 0.71 | 0.62 | |
| 4 | TheCowKing (▲1) | Test1 | 0.55 | 0.64 | 0.59 | 0.60 |
| | | Test2 | 0.56 | 0.67 | 0.60 | |
| 5 | goedleio (▼4) | Test1 | 0.55 | 0.60 | 0.57 | 0.58 |
| | | Test2 | 0.58 | 0.62 | 0.60 | |
| 6 | MNDS (▲2) | Test1 | 0.51 | 0.62 | 0.55 | 0.56 |
| | | Test2 | 0.51 | 0.62 | 0.56 | |
| 7 | DTND | Test1 | 0.51 | 0.49 | 0.49 | 0.53 |
| | | Test2 | 0.50 | 0.72 | 0.58 | |
| 8 | IISLABSKKU (▼2) | Test1 | 0.55 | 0.58 | 0.56 | 0.52 |
| | | Test2 | 0.72 | 0.37 | 0.48 | |
| 9 | suya (▲1) | Test1 | 0.50 | 0.40 | 0.44 | 0.42 |
| | | Test2 | 0.38 | 0.44 | 0.40 | |
| 10 | YK (▲2) | Test1 | 0.63 | 0.40 | 0.49 | 0.39 |
| | | Test2 | 0.64 | 0.22 | 0.33 | |
| 11 | GoAlone | Test1 | 0.29 | 0.85 | 0.42 | 0.35 |
| | | Test2 | 0.31 | 0.31 | 0.31 | |
| 12 | NoJam (▲1) | Test1 | 0.31 | 0.30 | 0.30 | 0.30 |
| | | Test2 | 0.31 | 0.31 | 0.31 | |
| 13 | Lessang (▼4) | Test1 | 0.30 | 0.29 | 0.29 | 0.29 |
| | | Test2 | 0.29 | 0.29 | 0.29 | |

*(b) Track 2*

| Rank | Team | Test1 score | Test2 score | Total score |
|---|---|---|---|---|
| 1 | **Yokozuna Data** ▲3) | **0.88** | **0.61** | **0.72** |
| 2 | IISLABSKKU (▲3) | 1.03 | 0.67 | 0.81 |
| 3 | UTU (▼2) | 0.92 | 0.89 | 0.91 |
| 4 | TripleS (▼1) | 0.95 | 0.89 | 0.92 |
| 5 | DTND (▼3) | 1.03 | 0.93 | 0.97 |

### B. Track 2

Although predicting the survival time of users carries more benefits for game companies, it is more difficult to produce accurate prediction models compared with predicting a simple classification as Track 1. Consequently, only five of the teams that participated in Track 1 also participated in Track 2. Yokozuna Data won this track as well, with a total score of 0.72.

---

[4] https://arxiv.org/abs/1802.02301

[5] https://cilab.sejong.ac.kr/gdmc2017/index.php/tutorial/

Unlike Track 1, performance was measured using RMSLE, in which a lower score is better. As with the results for Track 1, the test server results differed from those at the final outcome. The highest score from the test server was 0.41, whereas that of the final performance measure was around 0.72.

## V. METHODS USED BY PARTICIPANTS

### A. Overview

Throughout the competition, various methods were applied and tested on the game log data of Blade & Soul of NCSOFT. For Track 1, there was no overriding technique. All groups were evenly distributed in the rankings. This means that the combination of features, learning models, and feature selection is more important than what learning model is used to improve churn prediction accuracy.

Proper data pre-processing was highly important with regard to achieving high performance in the prediction. The winner, Yokozuna Data used the most various features among participants. They used daily features, entire period features, time-weighted features and statistical features. It is also interesting that Yokozuna Data used different algorithms for each of the test sets. In addition, deep learning appeared to be less popular than computer vision, speech analysis, and other engineering domains in the field of game data mining, as there were many participants who implemented tree-based ensemble classifiers (e.g., extra-tree classifiers and Random Forest) and logistic regression. These prediction techniques were classified into three major groups: neural network, tree-based, and linear model.

For Track 2, no entries with deep learning techniques were submitted; all of the submissions used either tree regression or linear models. The winner, Yokozuna Data, combined 900 trees for the regression task.

Those who performed well for Track 1 also excelled in Track 2, as the rankings among Track 2 participants were the same as those of Track 1, except for team IISLABSKKU. Although the rankings for Track 2 were somewhat similar to those for Track 1, the same was not the case for the usage of techniques, as no one implemented a neural network model to solve the problem for Track 2. Even team Yokozuna Data, which showed impressive performance for Track 1 using a neural network, chose not to apply neural network techniques to Track 2.

Now, we describe each method submitted to the competition, beginning with Yokozuna Data, the winner of the competition.

### B. Yokozuna Data (Winner of Track 1 and Track 2)

Here are summarized methods of the Yokozuna Data feature engineering process, which is based on previous works [27][28].

Yokozuna Data used two types of features: daily features and overall features. Daily features are the features which are aggregated the user's in-game activities by day, while overall features are calculated over the whole data period.

There were more than 3000 extracted features after the data preparation. Besides adopting all features, the following three feature selection methods were also tested: LSTM, auto encoder, feature value distribution and feature importance.

Multiple models were evaluated for both tracks. Additionally, different models were used for the two test datasets. The models that produced the best results – and that led us to win both tracks – were LSTM, extremely randomized trees and conditional inference trees. Parameters were adjusted through cross-validation.

*1) Binary Churn Prediction (Track 1)*

*1-1) LSTM*

A model combining an LSTM network with a deep neural network (DNN) was used for test set 1. First, a multilayer LSTM network was employed to process the time-series data and learn a single vector representation describing the time-series behavior. Then, this vector was merged with the output of a multilayer DNN that was trained on the static features. After merging, an additional layer was trained on top of the combined representation to get the final output layer predicting the binary result. In order to prevent overfitting, dropout [29] was used at every layer. Additionally, dropout was also applied to the input to perform random feature selection and further reduce overfitting. As there were many correlated features, selecting only a subset of the time-series and static features of the input prevents the model from depending too much on a single feature and improves generalization.

*1-2) Extremely Randomized Trees*

This technique provided better prediction results for the test set 2. After parameter tuning, 50 sample trees were selected and the minimum number of samples required to split an internal node was set to 50. Yokozuna Data used a splitting criterion based on the Gini impurity.

*2) Survival Time Analysis (Track 2)*

*2-1) Conditional Inference Survival Ensembles*

For both tests of the survival track, the best results were obtained with a censoring approach, using conditional inference survival ensembles. The final parameter selection was performed using 900 selected unbiased trees and subsampling without replacement. At each node of the trees, 30 input features were randomly selected from the total input sample. The stopping criteria employed was based on univariate p-values. For more details please check [27].

### C. UTU (2nd Place in Track 1 and 3rd Place in Track 2)

UTU used various features that represented activities such as availability, playing probability, message rate, session lengths, and entry time. These features were split into three features of overall, last week, and change measure. In addition, these activity features were extended by smart features of essentially previous activity (total experience, maximum experience, current experience, rating, money, etc.), whereby "smart" means that they reverse-engineered how the experience accumulation worked with regard to 'previous activity' measurement.

For classification, UTU used simple logistic regression, with some regularization. They did not attempt to use the cross-validated 'optimal settings' for parameters because they were unable to find any differences. Moreover, because of a covariate shift, they concluded that it would not necessarily be an optimal setting. Feature selection based on the basic L1-norm seemed useful in Test Set 2.

For regression, UTU used ridge regression. To linearize the features for both submissions, they used a feature transformation based on quantiles.

*D. TripleS (3rd place in Track 1 and 4th Place in Track 2)*

TripleS used play time (referring to how long the user was connected to the game), level and mastery level, the sum of mastery experience, log count, and specific log ID count as features. All features were extracted on a weekly basis. In addition, they calculated the coefficient of variance and first-order and quadratic functions of the abovementioned features and added them as features only for active players.

After preprocessing the features (normalizing and scaling), TripleS used an ensemble method, such as Random Forest, for Tracks 1 and 2. Moreover, to obtain the best result, the parameters for Random Forest were optimized. All of the results of Tracks 1 and 2 were validated by five-fold cross-validation with a training dataset.

*E. IISLABSKKU (8th Place in Track 1 and 2nd place in Track 2)*

IISLABSKKU created 1639 features through feature engineering. They calculated the feature's importance through Random Forest and used xgboost to calculate the final results for the 100 critical features.

*F. TheCowKing (4th Place in Track 1)*

TheCowKing used log counts on a weekly basis. Log ID and actor level were used as features. The main learning model of TheCowKing was LightGBM[6].

*G. goedle.io (5th Place in Track 1)*

*1) Feature engineering*

Goedle.io transformed the events for each player into an event-based format. Additional information, such as an event identifier (e.g., to differentiate a kill of a PC vs. NPC) or an event value (e.g., to track the amount of money spent), can be added to each event. While not all events provide meaningful identifiers or values, they added to roughly a third of the events an identifier, value, or both. Sometimes, more than one identifier or value is possible. In such cases, they duplicated the event and set the fields accordingly.

Many of the features used were initially inspired by [30][31]. Over the past years, they have added numerous additional features to their toolbox. The features can be categorized into different buckets:

- Basic activity: measuring basic activity such as the current absence time of a player, the average playtime of player, the number of days a player has been active, or the number of sessions.
- Counts: counting the number of times an event or an event-identifier combination occurs.
- Values: different mathematical operations applied to the event values, e.g., sum, mean, max, min, etc. (i.e., to the statistic features).
- Timings: timing between events to detect recurring patterns, e.g., a slowly decreasing retention.
- Frequencies: a player's activity can be transformed from a time-series into the frequency domain. Now the strongest recurring frequency of a player can be estimated and used as a feature.
- Curve Fitting: as described in [30], curve fitting can be applied to time series data. Parameters such as a positive slope of the fit suggest that the interest in the game is increasing.

Based on these features, datasets for training and testing sets generated. This resulted in almost 600 features for each player. But with only 4,000 players in the training dataset, one has to be careful with too many dimensions. Depending on the algorithm, feature selection and regularization are not only helpful but necessary. Additionally, they applied the outlier detection to the training dataset. Outlier detection helps to remove noise from the raw data and to exclude misleading data introduced by malicious players or bots. The outlier detection removed 0.5% of players in the training dataset.

*2) Modeling*

Goedle.io evaluated a variety of algorithms for each prediction problem. Some algorithms handle raw datasets quite well, e.g., tree-based algorithms, but other algorithms strongly benefit from a scaling the data. For that reason, they apply a simple scaler to the datasets.

To find the best performing algorithm including features, preprocessing steps, and parameters, they iteratively test different combinations. To evaluate algorithms and to measure the improvements due to the optimization, they used a 5-fold cross validation based on the F1-score. Goedle.io selected the best algorithm among different ones including: Logistic Regression (LR), Voted Perceptron (VP), Decision Trees (DT), Random Forests (RF), Gradient Tree Boosting (GTB), Support Vector Machines (SVM), and Artificial Neural Networks (ANN).

---

[6] https://github.com/Microsoft/LightGBM

## H. MNDS (6th Place in Track 1)

MNDS encoded the time-series data into image pixels, user variables were mapped to the x-axis of the image, and the game-use period (8 weeks) was linearly mapped to the y-axis of the image. The features were composed of 13 input variables, obtained by compressing 20 variables according to the order of variable importance provided from Random Forest model. The process was repeated, in which similar variables were removed. Ultimately, the image pixels per user were composed of $13 \times 56$ pixels. It's based on the work on tiled convolutional neural networks [32].

In addition, since the behavior occurring near the eighth week affects customer churn as opposed to the behavior of the first week, a distance-weighted coefficient $W_i$ is assigned to the reciprocal of the time 'distance' to give a weight for each day:

$$W_i = \frac{1}{d(t_q, t_i)} \quad (3)$$

$t_q$ : Date of q (query) point
$t_i$ : Last date + 1
$d(t_q, t_i)$ : Distance between two dates

Next, to convert the weighted time-series data into image pixels using a Gramian Angular Field (GAF), min-max normalization was performed with a value between $[-1, 1]$, according to Eq. (4):

$$\tilde{x}^{(i)} = \frac{(x^{(i)} - \max(X)) + (x^{(i)} - \min(X))}{\max(X) - \min(X)} \quad (4)$$

The next step of the GAF is to represent the value of the normalized value $\tilde{X}$ in the coordinates of the complex plane, where the angle is expressed by the input value and the radius $i$ is expressed by the time axis.

$$\emptyset = \arccos(\tilde{x}^{(i)}), \quad -1 \leq \tilde{x}^{(i)} \leq 1, \quad \tilde{x}^{(i)} \in \tilde{X} \quad (5)$$
$$r = \frac{i}{N}, \quad i \in N$$

Finally, the GAF is defined below, and the G matrix is encoded as an image.

$$G = \begin{bmatrix} \cos(\emptyset_1 + \emptyset_1) & \cdots & \cos(\emptyset_1 + \emptyset_n) \\ \vdots & \ddots & \vdots \\ \cos(\emptyset_n + \emptyset_1) & \cdots & \cos(\emptyset_n + \emptyset_n) \end{bmatrix} \quad (6)$$

The GAF provides a way to preserve the temporal dependence, since time increases as the position moves from top-left to bottom-right. The GAF contains temporal correlations, as the G matrix represents the relative correlation by superposition of directions with respect to the time interval k.

MNDS then applied the modified model of Inception-V3 [33] into image pixel data for churn prediction.

## I. DTND (7th Place in Track 1 and 5th Place in Track 2)

DTND supposed that the data were noisy. Hence, GLM and Elastic-net were applied to the learning model. The feature set included days played, action counts, action type diversity, and file size. None of the features had a negative correlation. With vanilla GLM, owing to offset overloaded features, negative correlation factors exist. In addition, they predicted non-churn instead of churn.

## VI. DISCUSSION: POST-CONFERENCE ANALYSIS

### A. Difficulty of the competition problems

Churn prediction of only loyal users differed considerably from that based on all user types. When conducting churn prediction targeting all users, the task becomes relatively straightforward as the majority of users show obvious churning signals. On the other hand, loyal users seldom churn or their churn is often due to out-of-game-related issues, which hinders churn prediction based on in-game activity data. Our comparison of performance between predicting churn of all users and that for loyal users only confirmed that the performance differed significantly with respect to prediction group targets.

Two experiments were conducted to compare prediction performance regarding churn of all users and of loyal users only. For the first experiment, both training and evaluation sets were created with all user data, whereas for the second experiment, the training and evaluation sets were created using only data from loyal users. Each experiment was trained on data of 3,000 users and was evaluated using a dataset consisting of 1,400 users' data.

Five machine learning algorithms, Random Forest, Logistic Regression, Extra Gradient Boosting (XGB), Generalized Boosting Model (GBM), and Conditional Inference Tree, were applied to churn prediction of all users and loyal users only. When each machine learning algorithm was trained and evaluated on a dataset with all users, the F1 score ranged approximately from 0.6 to 0.72. When the same procedure was repeated with a dataset of only loyal users, the F1 score ranged from 0.39 to 0.53, showing a significant drop in overall predictive performance (Figure 8).

Owing to the emphasis on chronological robustness, we expected that imposing a time difference between training and test sets would lower the performance of participants compared to other churn prediction work, as confirmed by a comparative experiment of churn prediction of users. The experiment was conducted using two evaluation sets where one was created with data from the same period as the training set and the other of data from periods two months after. The results revealed much lower performance by predicting the churn of users from different periods.

Using the same five machine learning algorithms, prediction models were trained on user data from Nov. 1, 2017, to Nov. 30, 2017. Performance was measured using a test set containing data of concurrent yet different users, yielding a mean F1 score of approximately 0.45. The same models were used to predict the churn of different users two months later – from Jan 1, 2018, to Jan 31, 2018 – yielding a significantly lower F1 score ranging from 0.04 to 0.3 (Figure 9).

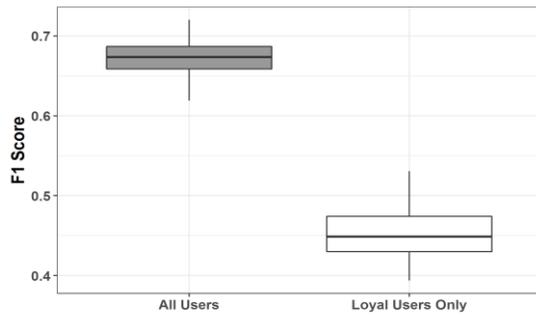

*Figure 8. F1 score distribution resulting from a comparative experiment of churn prediction using five machine learning algorithms. Higher predictive performance was achieved when training and evaluation were conducted using data with all users (left) compared to when the same process was applied to a dataset comprising only loyal users (right)*

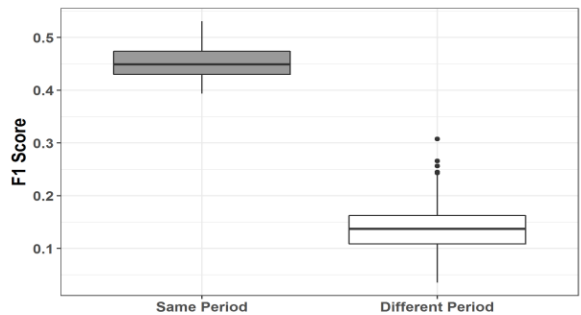

*Figure 9. Performing churn prediction with concurrent training and evaluation data (left) shows more accurate prediction compared with performing the same task with training and evaluation data created from different periods (right)*

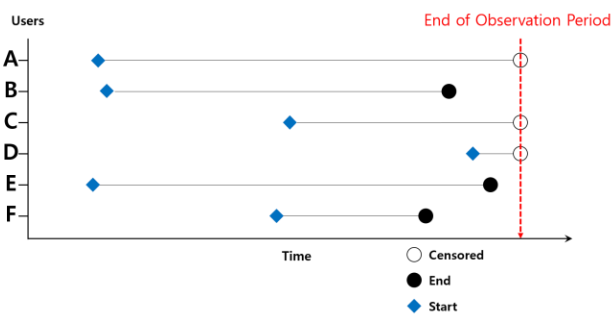

*Figure 10. Data censoring (right censoring)*

Based on these experimental results, we expected the participants' prediction performances to be lacking compared with those of previous studies.

### B. Right censoring issue in Track 2

For any survival analysis problem, the exact survival time may be incalculable due to censoring, in which some data are only partially known due to actual constraints, as opposed to model design. The most common type of censoring is right censoring, where survival time is at least as long as the observation period, as shown in Figure 10. Owing to the censoring nature of survival analysis, we predicted that employing a method that correctly incorporates both censored and uncensored data would be crucial. Consequently, it was not surprising to find that Yokozuna Data, the only group to consider the censoring nature of the problem explicitly and implement an effective method, namely, a conditional inference survival ensemble, exhibited superior performance and won the track.

Contestants also had to consider that the performance assessments via a test server and the final results were performed through calculation using different censoring points. The final evaluation was made based on the survival time calculated on July 31, 2017, whereas the survival time for the test server evaluation was calculated based on survival time by March 28, 2017. We expect that the consideration of such different censoring points between the test server and final evaluation would be crucial to prevent overfitting to the test server and attain accurate predictions; such expectations coincided with the final results. Yokozuna Data and IISLABSKKU were the only teams that not only reached the top places—1st and 2nd—but also improved their final standings compared with test server standings; they also appeared to be the only contestants who considered the different censoring points.

When the distributions of survival time of each team for the second test set were compared, there was a significant difference between the survival time prediction of Yokozuna Data and IISLABSKKU (Group A), whose final standing improved greatly compared with the standings from the test server results and the other participants (Group B) whose standing deteriorated.

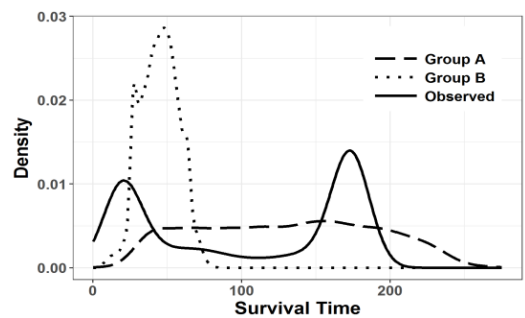

*Figure 11. Distribution of survival time for Test 2 compared among the prediction results of Groups A and B and observed survival times.*

As shown in Figure 11, the survival time predicted by Group A had a wider spread, which indicated that Group A predicted users to survive far beyond the test server and final evaluation censoring point. In contrast, Group B's prediction did not seem to consider survival beyond the censoring point of the test server. Because Group B predicted significantly shorter survival times, we suspect that their results were over-fitted to the test server results.

Furthermore, our performance measurement for survival time prediction should be improved in the future. Figure 12 shows survival curves of participants' prediction result for test set 1 and test set 2. According to the Figure 12, we suspect that RMSLE tends to benefit from overestimating survival time. We think time-dependent ROC curve can be a good alternative for measuring a performance of survival time prediction [34].

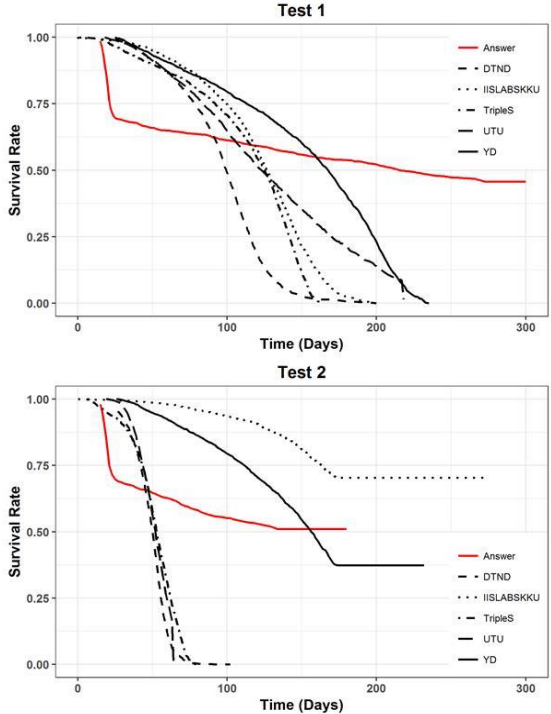

*Figure 12. Survival curves of Test set 1 and 2 in Track 2.*

*C. Future works to improve the competition*

To run the competition, we provided training and test datasets (without labels) and allowed the participants to build models. This means that they could use the test data for unsupervised learning tasks to build more generalized models. It is important to measure the progress of the participants and encourage them to compete with each other. For that purpose, we supported a test-server which return test accuracy on 10% of datasets (sampled from the test sets). Participants used the test server to monitor their model's progress and their current position on the leaderboard of the server. Although the test server is a useful tool for competition, there are several issues to solve for future game data mining competitions.

- For the competition organizers, it is important to run the test server securely. For example, it must not be possible to get the ground truth of the test sets used in the test server. In the ImageNet competition, for example, there was a case in which the participants used the test server illegally by opening multiple accounts.
- For the participants, it is important to provide useful information from the test server to enable the best performance. However, it is challenging to run the competition server properly while gamers on the test dataset are still playing the Blade & Soul game. In Track 2 (survival analysis), the ground truth of the test sets was not determined before the final evaluation of the entries because the gamers on the test log data still played the games and their ground truth value (survival weeks) changed, up until the last evaluation.

Compared with other game AI competitions, the number of participants in this game data mining competition was not small; however, future efforts should attempt to attract more data science researchers to the competition. In terms of participant diversity, the competition should attempt to attract participants from many different countries (e.g., this year, South Korea was the dominant country in the competition). This requires the use of various advertisement venues to attract a more diverse group of international participants. This year, participants used the test server implemented by our team; however, well-known data competition platforms such as Kaggle (https://www.kaggle.com/) may help the competition promote.

VII. CONCLUSIONS

In this paper, we propose a competition framework for game data mining using commercial game log data. The goal of the competition was very different from other types of game AI competitions that targeted strong or human-like AI players and content generators. Here, the goal of the game data mining competition was to build an accurate predictor trained using game log data. The first track focused on the classification problem to predict churn or no churn as a binary decision task. The second task involved solving a regression problem to predict the number of weeks that the gamer would survive in the future. From a practical perspective, the second task is more desirable for the game industry; however, it is more difficult to make an accurate prediction for this.

Furthermore, we designed the competition problem in consideration of the various practical issues in live game. First, a much longer time span was given between the training data and prediction window, reflecting the minimum time required to execute churn prevention strategies. Second, test sets include a change of business model to drive concept drift issue. Third, we provided only log data of loyal users for the competition. According to our experiments, they tend to be more difficult to predict churn than others, however they are more valuable in business.

Throughout the competition, various methods were applied and tested on the game log data of Blade & Soul of NCSOFT. A total of 13 teams participated in track I and 5 teams in track II. Yokozuna Data won the both.

Our approach marks the first step in opening a public dataset for game data mining research and provides useful benchmarking tools for measuring progress in the field. For example, Kummer *et al.*, recently reported the use of commitment features using the benchmarking dataset [35]. Although we limited the competition's goal to predict players' churning behavior and number of surviving weeks, another purpose of our game data mining competition is to promote the research of game data mining by providing commercial game logs to the public.

ACKNOWLEDGEMENTS

We'd like to express great thanks to the all contributors. All the data used in this competition is freely available through GDMC 2017 homepage (https://cilab.sejong.ac.kr/gdmc2017/) and google groups (https://groups.google.com/d/forum/gdmc2017). This research is partially supported by Basic Science Research Program through the National Research Foundation of Korea(NRF) funded by the Ministry of Science, ICT & Future Planning (2017R1A2B4002164). This research is partially supported by Ministry of Culture, Sports and Tourism (MCST) and Korea Creative Content Agency(KOCCA) in the Culture Technology(CT) Research & Development Program 2017. This work is partially supported by the European Commission under grant agreement number 731900 - ENVISAGE. Supplementary material file is available from https://arxiv.org/abs/1802.02301

Supplementary Material

The detailed information of participants

*Table A. Summary of participants*

| Team | # of Members | Affiliation | Country |
|---|---|---|---|
| DTND | 3 | - | Korea |
| GoAlone | 1 | Yonsei Univ. | Korea |
| goedleio | 2 | goedle.io GmbH | Germany |
| IISLABSKKU | 3 | Sungkyunkwan Univ. | Korea |
| Lessang | 2 | Yonsei Univ. | Korea |
| MNDS | 3 | Yonsei Univ. | Korea |
| NoJam | 3 | Yonsei Univ. | Korea |
| suya | 1 | Yonsei Univ. | Korea |
| TheCowKing | 2 | KAIST | Korea |
| TripleS | 3 | - | Korea |
| UTU | 4 | Univ. of Turku | Finland |
| Yokozuna Data | 4 | Silicon Studio | Japan |
| YK | 1 | Yonsei Univ. | Korea |

*Table B. Summary of features used by participants*

| Rank | Team | sampling period | entire period features | time weight features | Statistic features | Variable selection |
|---|---|---|---|---|---|---|
| 1 | Yokozuna Data | daily | O | O | O | O |
| 2 | UTU | ? | O | O | O | O |
| 3 | TripleS | weekly | X | X | O | O |
| 4 | TheCowKing | weekly | X | X | O | X |
| 5 | goedleio | daily | X | X | X | X |
| 6 | MNDS | daily | X | O | X | O |
| 7 | DTND | ? | O | X | X | X |
| 8 | IISLABSKKU | ? | ? | ? | ? | O |
| 9 | suya | X | O | X | X | X |
| 10 | YK | X | O | O | X | O |
| 11 | GoAlone | X | O | X | X | X |
| 12 | NoJam | ? | ? | ? | ? | ? |
| 13 | Lessang | daily | X | X | X | O |

Table C. Techniques used by each team in Track I (LSTM = Long-Short-Term Memory, DNN = Deep Neural Network).

*(a) Track 1*

| Rank | Team | Techniques |
|---|---|---|
| 1 | Yokozuna Data | LSTM+DNN, Extra-Trees Classifier |
| 2 | UTU | Logistic Regression |
| 3 | TripleS | Random Forest |
| 4 | TheCowKing | LightGBM (Light Gradient Boosting Machine) |
| 5 | goedleio | Feed Forward Neural Network, Random Forest |
| 6 | MNDS | Deep Neural Network |
| 7 | DTND | Generalized Linear Model |
| 8 | IISLABSKKU | Tree Boosting |
| 9 | suya | Deep Neural Network |

| 10 | YK | Logistic Regression |
|----|----|---------------------|
| 11 | GoAlone | Logistic Regression |
| 12 | NoJam | Decision Tree |
| 13 | Lessang | Deep Neural Network |

*(b) Track 2*

| Rank | Team | Techniques |
|------|------|------------|
| 1 | Yokozuna Data | Ensemble of Conditional Inference Trees (# of Trees = 900) |
| 2 | IISLABSKKU | Tree Boosting |
| 3 | UTU | Linear Regression |
| 4 | TripleS | Ensemble Tree Method |
| 5 | DTND | Generalized Linear Model |

| Neural Net | Tree Approach | Linear Models |

The statistics of each grade mentioned in III-B

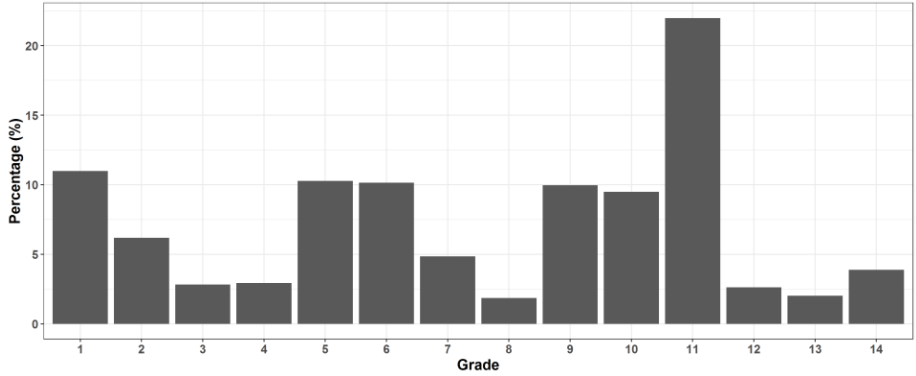

*Figure A. User population distribution among different user grades during the training data period*

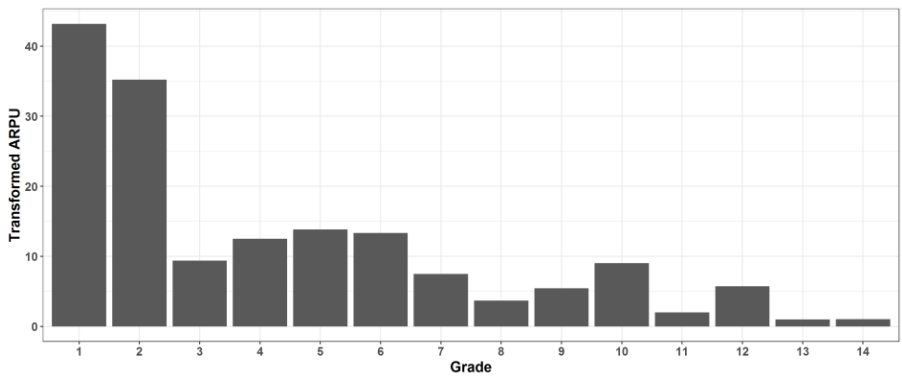

*Figure B. Transformed Average Revenue Per User (ARPU) for each grade during the training data period.*

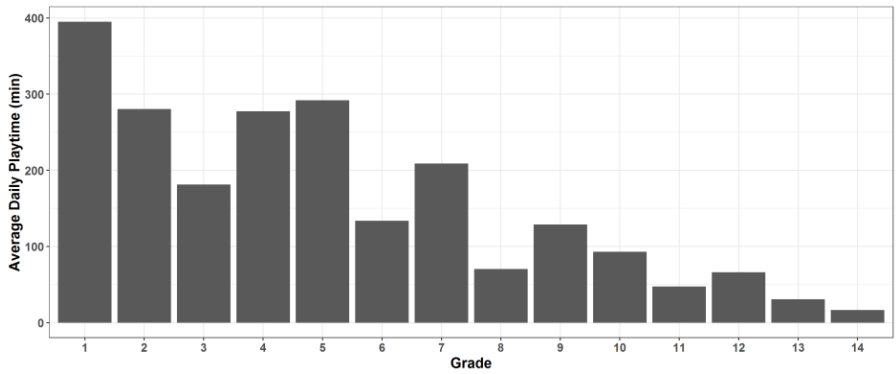

*Figure C. Average daily playtime for each grade during the training data period.*

*The detailed materials of Yokozuna Data*

*Table D. The summary of features which are used by Yokozuna Data*

| Type | Feature | Description |
| --- | --- | --- |
| Daily | Action | Number of times each action was performed per day. Some actions were grouped, and then their counts were summed up. For example, every time a player created, joined, or invited other players to a guild was counted as a "Guild" feature. |
| | Session | Number of sessions per day. The session was considered to be over once the player had been inactive for more than 15 minutes. |
| | Playtime | Total playtime (summing all sessions) per day. |
| | Level | Daily last level and number of level-ups per day. |
| | Target level | Daily highest battle-target level. If the player did not engage in any battle on a certain day, the target level of that day was set to the value of the previous day. |
| | Number of actors | Number of actors played per day. |
| | In-game money | Total amount of money earned and spent per day. |
| | Equipment | Daily equipment score. The higher this score, the better the equipment owned by the player. If there was no equipment score logged on a certain day, the equipment score of that day was set to the value of the previous day. |
| Overall | Statistics | Statistics, such as the mean, standard deviation, and sum of all the time-series features. For example, the total amount of money the player got during the data period or the standard deviation of the daily highest battle-target level. |
| | Relation bewween the first and last days | Differences in the behavior of a player between the first and last days in the data period were calculated to measure changes over time. For example, the difference between the total number of level-ups in the first three days and the last three days. |
| | Actors information | Number of actors usually played. |
| | Loyalty index | Percentage of days in which the player was connected between their first and last connection. |
| | Guilds | Total number of guilds joined. |
| | Information of days of weeks | Distribution of actions over different days of weeks. |

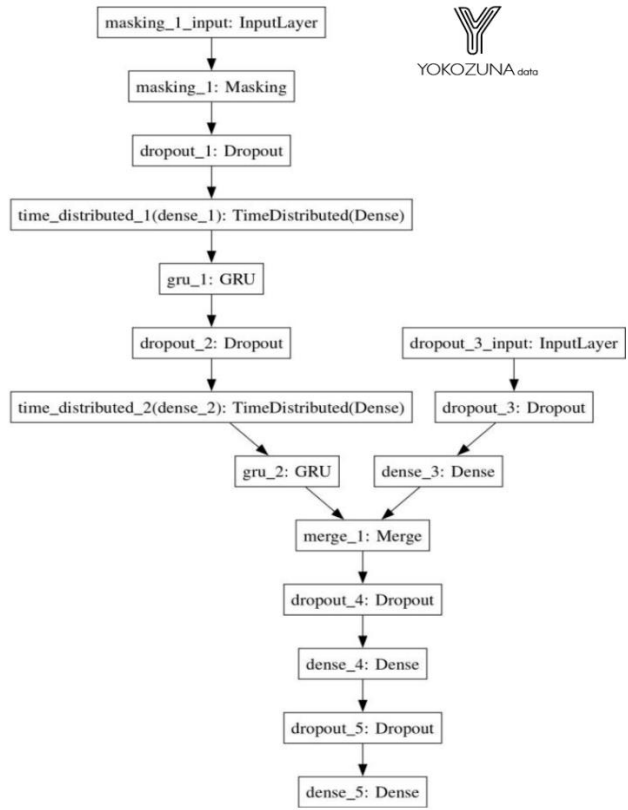

*Figure D. Neural network structure for the test set 1 in the binary case (track 1). The left part is an LSTM network with the time-series data as inputs. The right part is a DNN with the static features presented as inputs. The bottom part is a DNN which merges the outputs of these two networks and provides the final prediction result*

*The detailed materials of geodle.io*

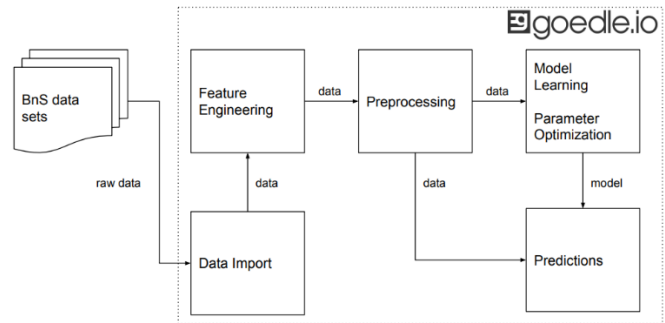

*Figure E. goedle.io's process pipeline from raw data to churn prediction*

I. Time performance:

- Frequencies A player's activity can be transformed from a time series into the frequency domain. Now the strongest recurring frequency of a player can be estimated and used as a feature.
- DT, RF, GTB: Learning and prediction take seconds.
- SVMs: Learning and prediction also run in seconds. Learning is for certain parameter combinations computationally more challenging but still runs in seconds or a few minutes at most. We removed more demanding configurations from the setting.
- ANNs: Learning and prediction are possible in seconds with GPUs. The required time quickly increases with the number and size of hidden layers. We therefore limited ourselves to small networks with four hidden layers or less.
- All of the experiments, except for the ANNs, were conducted on a server with two Intel Xeon processors with 8 cores at 2.6GHz. The ANNs were computed on an Nvidia GTX 970 with native CUDA support.

II. Insights

- For DT, RF, and GTB, features based on counts, values, timings, and fits have shown to be often almost equally important. With counts of very basic events such as "EnterWorld" being quite prominent at the top of the lists when sorted by feature importance.
- For SVMs, features based on counts are comparably more prominent among the most important features. Those counts present a strong indicator for churn — or loyalty in the case of an opposite sign of the weight.
- Relational features such as the centrality of players obtained from the social graph also show to be important in many experiments with different algorithms.
- Opposed to other work and experiments seen at goedle.io, the current absence time does not discriminate churners from non-churners well in Blade & Soul. We assume that the most obvious churners, i.e., players with high current absence time and a low number of events, have been previously removed from the data. This would also explain the overall low number of churners in the dataset. The dataset contains 30% churners, which is quite low when compared to other mobile games and apps.
- Activeness of party members was not among the top features. This is again in contrast to our expectations and previous experience. One possible explanation for this observation is the fact that the data for the competition only depicts a subset of the entire social network. While the training dataset contains only 4,000 users, our party graph lists more than 32,000 identifiers of users. I.e., the activity of only roughly 10% of the entire network is observed.
- Often, only roughly 3% of players spend money in free-to-play games. Therefore, it is important to carefully craft an outlier detection that does not inaccurately remove high spending players as they might look like outliers or fraud.
- A single algorithm does not fit all problems: Often, there is not one algorithm that always works the best independent of the features or the distribution of the target variable. Some algorithms are more tolerant towards the distribution of the data and others can be trained more easily with unbalanced data. Additionally, the data contains a concept shift where tree-based algorithms worked well.
- Deep Learning is not superior to our platform. However, this should be taken with a grain of salt. We did not exploit the entire power and expressiveness of modern ANNs and Deep Learning. The structure of our ANNs is based on simple configurations and we suggest to testing Recurrent Neural Networks in the future.

*The detailed materials of MNDS*

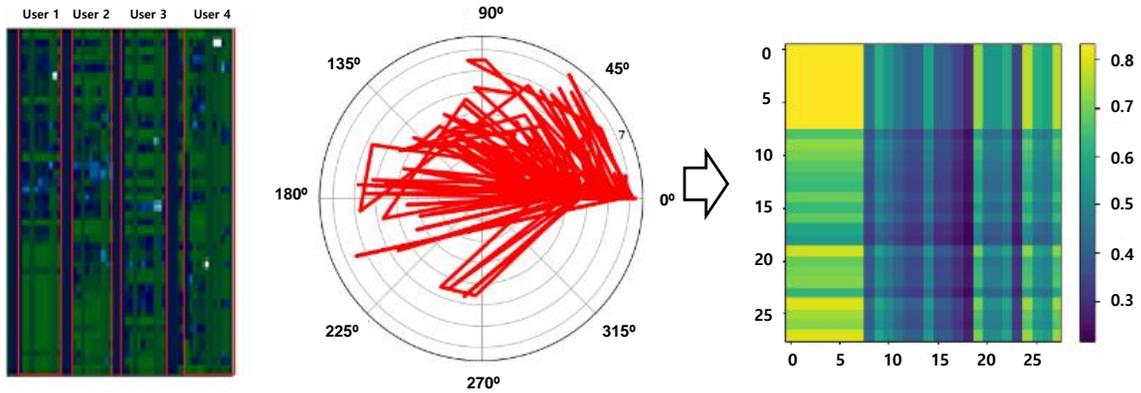

*Figure F. Encoding user log data (a) time (8 weeks) and feature are composed of row and column, which is the log data converted by user per image pixel. In the case of (b), the GAF method is used to transform the log data of the user into the polar coordinate system, and the user log image pixels are generated by applying the GAF method.*

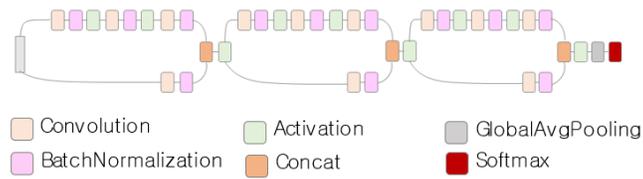

*Figure D. MNDS model*